\documentclass[aps,showpacs,nofootinbib]{revtex4}

\usepackage{graphicx}
\usepackage{graphics}
\usepackage{amssymb}
\usepackage{amsmath}
\usepackage{bm}

\newcommand{\sT}{{\scriptscriptstyle T}}
\newcommand{\g}{\gamma}

\newcommand{\be}{\begin{equation}}
\newcommand{\ee}{\end{equation}}
\newcommand{\bea}{\begin{eqnarray}}
\newcommand{\eea}{\end{eqnarray}}
\newcommand{\beal}{\begin{align}}
\newcommand{\eal}{\end{align}}
\newcommand{\bespl}{\begin{split}}
\newcommand{\espl}{\end{split}}
\newcommand{\nn}{\nonumber}
\newcommand{\nslash}{\kern 0.2 em n\kern -0.50em /}
\newcommand{\kslash}{\kern 0.2 em k\kern -0.45em /}
\newcommand{\pslash}{\kern 0.2 em p\kern -0.50em /}
\newcommand{\Sslash}{\kern 0.2 em S\kern -0.50em /}
\newcommand{\Pslash}{\kern 0.2 em P\kern -0.50em /}
\newcommand{\Rslash}{\kern 0.2 em R\kern -0.50em /}

\begin{document}
\title{
Monte Carlo simulation of single spin asymmetries in pion-proton collisions}

\author{Andrea Bianconi}
\email{andrea.bianconi@bs.infn.it}
\affiliation{Dipartimento di Chimica e Fisica per l'Ingegneria e per i 
Materiali, Universit\`a di Brescia, I-25123 Brescia, Italy, and\\
Istituto Nazionale di Fisica Nucleare, Sezione di Pavia, I-27100 Pavia, Italy}

\author{Marco Radici}
\email{marco.radici@pv.infn.it}
\affiliation{Dipartimento di Fisica Nucleare e Teorica, Universit\`{a} di 
Pavia, and\\
Istituto Nazionale di Fisica Nucleare, Sezione di Pavia, I-27100 Pavia, Italy}

\begin{abstract}
We present Monte Carlo simulations of both the Sivers and the Boer-Mulders effects in 
the polarized Drell-Yan $\pi^\pm p^\uparrow \to \mu^+ \mu^- X$ process at the 
center-of-mass energy $\sqrt{s} \sim 14$ GeV reachable at COMPASS with pion beams of
energy 100 GeV. For the Sivers effect, we adopt two different parametrizations for the 
Sivers function to explore the statistical accuracy required to extract unambiguous
information on this parton density. In particular, we verify the possibility of 
checking its predicted sign change between Semi-Inclusive Deep-Inelastic Scattering 
(SIDIS) and Drell-Yan processes, a crucial test of nonperturbative QCD. For the
Boer-Mulders effect, because of the lack of parametrizations we can make only guesses.
The goal is to explore the possibility of extracting information on the transversity 
distribution, the missing piece necessary to complete the knowledge of the nucleon spin 
structure at leading twist, and the Boer-Mulders function, which is related to the 
long-standing problem of the violation of the Lam-Tung sum rule in the unpolarized 
Drell-Yan cross section.
\end{abstract}

\pacs{13.75.Gx,13.88+e,13.85.Qk}

\maketitle

\section{Introduction}
\label{sec:intro}

The recent measurement of Single-Spin Asymmetries (SSA) in semi-inclusive $l
p^\uparrow \to l'\pi X$ Deep-Inelastic Scattering (SIDIS) on transversely
polarized hadronic 
targets~\cite{Airapetian:2004tw,Diefenthaler:2005gx,Avakian:2005ps,Alexakhin:2005iw}, 
has renewed the interest in the problem of describing the spin structure of 
hadrons within Quantum Chromo-Dynamics (QCD)~\cite{cerncourier}, and has 
stimulated since then a large production of phenomenological and theoretical 
papers. Experimental evidence of large SSA in hadron-hadron collisions was well 
known since many years~\cite{Bunce:1976yb,Adams:1991cs}, but it has never been 
consistently explained in the context of perturbative QCD in the collinear 
massless approximation~\cite{Kane:1978nd}. The idea of going beyond the 
collinear approximation opened new perspectives about the possibility of 
explaining these SSA in terms of intrinsic transverse motion of partons inside 
hadrons, and of correlations between such intrinsic transverse momenta and 
transverse spin degrees of freedom. The most popular examples are the 
Sivers~\cite{Sivers:1990cc} and the Collins~\cite{Collins:1993kk} effects. In 
the former case, an asymmetric azimuthal distribution of detected hadrons (with 
respect to the normal to the production plane) is obtained from the 
nonperturbative correlation ${\bm p}_\sT \times {\bm P}\cdot {\bm S}_\sT$, where 
${\bm p}_\sT$ is the intrinsic transverse momentum of an unpolarized parton 
inside a target hadron with momentum ${\bm P}$ and transverse polarization 
${\bm S}_\sT$. In the latter case, the asymmetry is obtained from the 
correlation ${\bm k} \times {\bm P}_{h\sT} \cdot {\bm s}_\sT$, where a parton 
with momentum ${\bm k}$ and transverse polarization ${\bm s}_\sT$ fragments into 
an unpolarized hadron with transverse momentum ${\bm P}_{h\sT}$. In both cases, 
the sizes of the effects are represented by new Transverse-Momentum Dependent 
(TMD) partonic functions, the socalled Sivers and Collins functions, 
respectively. 

However, SSA data in hadronic collisions have been collected so far typically for
semi-inclusive $pp^{(\uparrow )}\to h^{(\uparrow )} X$ processes, where the
factorization proof is complicated by higher-twist correlators~\cite{Qiu:1991pp} 
and the power-suppressed asymmetry can be produced by several (overlapping) 
mechanisms. On the contrary, the theoretical situation of the SIDIS measurements 
is more transparent. On the basis of a suitable factorization 
theorem~\cite{Ji:2004wu,Collins:2004nx}, the cross section at leading twist 
contains convolutions involving separately the Sivers and Collins functions with 
different azimuthal dependences, $\sin (\phi - \phi_S)$ and $\sin (\phi + 
\phi_S)$, respectively, where $\phi, \phi_S,$ are the azimuthal angles of the 
produced hadron and of the target polarization with respect to the axis defined 
by the virtual photon~\cite{Boer:1998nt}. According to the extracted azimuthal 
dependence, the measured SSA can then be clearly related to one effect or the 
other~\cite{Airapetian:2004tw,Diefenthaler:2005gx}.

Similarly, in the Drell-Yan process $H_1 H_2^\uparrow \to l^+ l^- X$ the cross
section displays at leading twist two terms weighted by $\sin (\phi - \phi_S)$
and $\sin (\phi + \phi_S)$, where now $\phi, \phi_S,$ are the azimuthal 
orientations of the final lepton plane and of the hadron polarization with 
respect to the reaction plane~\cite{Boer:1999mm}. Adopting the notations 
recommended in Ref.~\cite{Bacchetta:2004jz}, the first one involves the 
convolution of the Sivers function $f_{1\sT}^\perp$ with the standard 
unpolarized parton distribution $f_1$. The second one involves the transversity 
distribution $h_1$ and the Boer-Mulders function $h_1^\perp$, a TMD distribution 
which is most likely responsible for the violation of the Lam-Tung sum rule in 
the corresponding anomalous $\cos 2\phi$ asymmetry of the unpolarized Drell-Yan 
cross section~\cite{Boer:1999mm}. Hence, a simultaneous measurement of 
unpolarized and single-polarized Drell-Yan cross sections would allow to extract 
all the unknowns from data~\cite{Bianconi:2004wu,Bianconi:2005px}. Both $h_1$ 
and $h_1^\perp$ describe the distribution of transversely polarized partons; but 
the former applies to transversely polarized parent hadrons, while the latter to 
unpolarized ones. On an equal footing, $f_{1\sT}^\perp$ and $f_1$ describe 
distributions of unpolarized partons. The correlation between ${\bm p}_\sT$ and 
${\bm S}_\sT$ inside $f_{1\sT}^\perp$ is possible only for a nonvanishing 
orbital angular momentum of partons. Then, extraction of Sivers function from 
SIDIS and Drell-Yan data would allow to study the orbital motion and the spatial
distribution of hidden confined partons~\cite{Burkardt:2003je}, as well as to
test its peculiar universality property~\cite{Collins:2002kn}. 

In a series of previous papers, we performed numerical simulations of 
single-polarized Drell-Yan SSA for the $p p^\uparrow \to \mu^+ \mu^- 
X$~\cite{Bianconi:2005yj} and $\bar{p} p^\uparrow \to \mu^+ \mu^- 
X$~\cite{Bianconi:2004wu} processes. With proton beams, we considered collisions 
at $\sqrt{s}=200$ GeV in the kinematic conditions for the foreseen upgrade of 
RHIC (RHIC II). Even if in $pp$ collisions the nonvalence partonic contribution 
to the elementary annihilation is unavoidable (leading, in principle, to lower 
counting rates), still the kinematics selects a portion of phase space that 
emphasizes this contribution. The net result is that with a reasonable sample of 
Drell-Yan events the statistical accuracy allows to unambiguously extract the 
Sivers function from the corresponding $\sin (\phi -\phi_S)$ asymmetry, as well 
as to clearly test its predicted sign change with respect to the SIDIS 
asymmetry~\cite{Bianconi:2005yj}. In $\bar{p} p$ collisions, the cross section 
is dominated by the valence contribution to the annihilation of a parton (from 
$p$) and an antiparton (from $\bar{p}$); hence, in general it is not suppressed 
as in the previous case (for a quantitative check in our Monte Carlo, see Sec. 
IVB of Ref.~\cite{Bianconi:2005yj}). In Ref.~\cite{Bianconi:2004wu}, we selected 
antiproton beams of 15 GeV, as they could be produced at the High Energy Storage 
Ring (HESR) at GSI~\cite{Maggiora:2005cr,pax2}, and we simulated collisions at 
$\sqrt{s}\sim 14$ GeV in the socalled asymmetric collider mode. The goal was to 
explore the minimal conditions required for an unambiguous extraction of $h_1$ 
and $h_1^\perp$ from a combined analysis of the $\sin (\phi + \phi_S)$ and $\cos
2\phi$ asymmetries in the full (unpolarized + polarized) cross section.

Here, we will reconsider the same scenarios but for the $\pi^\pm p^\uparrow \to
\mu^+ \mu^- X$ process at the same $\sqrt{s}\sim 14$ GeV that can be reached at
COMPASS with pion beams of energy 100 GeV. As for $\bar{p}$ beams, the elementary
mechanism is dominated by the annihilation between valence partons (from $p$) and
valence antipartons (from $\pi$). Indeed, a large Sivers effect was predicted in
this context by usig the same Sivers function fitted to the measured $\sin (\phi 
-\phi_S)$ asymmetry in SIDIS~\cite{Collins:2005rq}. Taking advantage on the high 
statistics reachable with pions, in our Monte Carlo we simulate both $\sin (\phi 
\pm \phi_S)$ SSA in the Drell-Yan cross section. For the Sivers effect we use 
two parametrizations of $f_{1\sT}^\perp$: the one of 
Ref.~\cite{Anselmino:2005ea}, which was deduced by fitting the recent HERMES 
data for the $\sin (\phi - \phi_S)$ SSA~\cite{Diefenthaler:2005gx}; the one of 
Ref.~\cite{Bianconi:2005yj}, which is constrained by the recent RHIC data for 
the $pp^\uparrow \to \pi X$ process at higher energy~\cite{Adler:2005in}. For 
the Boer-Mulders effect, since there is no such abundance of data and fits, we 
follow Ref.~\cite{Boer:1999mm} to constrain $h_1^\perp$ by the azimuthal asymmetry of
the corresponding unpolarized Drell-Yan cross section (see also 
Ref.~\cite{Sissakian:2005yp} for a similar analysis). Then, we insert, as we did in 
Ref.~\cite{Bianconi:2004wu}, very different input test functions for $h_1$ in order to  
explore the sensitivity of the simulated SSA within the statistical accuracy. 

In Sec.~\ref{sec:mc}, we review the formalism and the details of the numerical 
simulation. In Sec.~\ref{sec:out}, we present and discuss our results. Finally, 
in Sec.~\ref{sec:end} we draw some conclusions.

\begin{figure}[h]
\centering
\includegraphics[width=7cm]{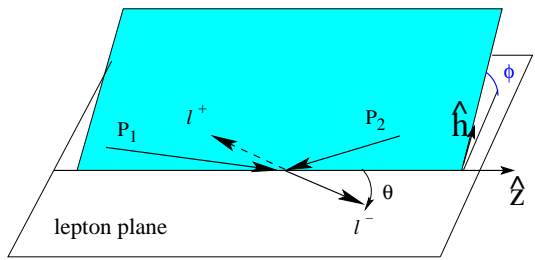}
\caption{The Collins-Soper frame.}
\label{fig:dyframe}
\end{figure}

\section{General framework for the numerical simulation}
\label{sec:mc}

In a Drell-Yan process, an antilepton-lepton pair with individual momenta $k_1$ 
and $k_2$ is produced from the collision of two hadrons with momentum $P_i$, mass 
$M_i$, and spin $S_i$, with $i=1,2$. The center-of-mass (c.m.) square energy 
available is $s=(P_1+P_2)^2$ and the invariant mass of the final lepton pair is 
given by the time-like momentum transfer $q^2 \equiv M^2 = (k_1 + k_2)^2$. If 
$M^2,s \rightarrow \infty$, while keeping the ratio $0\leq \tau = M^2/s \leq 1$ 
limited, a factorization theorem can be proven~\cite{Collins:1984kg} ensuring 
that the elementary mechanism proceeds through the annihilation of a parton and 
an antiparton with momenta $p_1$ and $p_2$, respectively, into a virtual photon
with time-like momentum $q^2$. If $P_1^+$ and $P_2^-$ are the dominant light-cone 
components of hadron momenta in this regime, then the partons are approximately 
collinear with the parent hadrons and carry the light-cone momentum fractions 
$0\leq x_1 = p_1^+ / P_1^+ , \; x_2 = p_2^- / P_2^- \leq 1$, with $q^+ = p_1^+, 
\; q^- = p_2^-$ by momentum conservation~\cite{Boer:1999mm}. The transverse
components ${\bm p}_{i\sT}$ of $p_i$ with respect to the direction defined by 
${\bm P}_i (i=1,2)$, are constrained again by the momentum conservation 
${\bm q}_\sT = {\bm p}_{1\sT} + {\bm p}_{2\sT}$, where ${\bm q}_\sT$ is the 
transverse momentum of the final lepton pair. If ${\bm q}_\sT \neq 0$ the 
annihilation direction is not known. Hence, it is convenient to select the 
socalled Collins-Soper frame~\cite{Collins:1977iv} described in 
Fig.~\ref{fig:dyframe}. The final lepton pair is detected in the solid angle 
$(\theta, \phi )$, where, in particular, $\phi$ (and all other azimuthal angles) 
is measured in a plane perpendicular to the indicated lepton plane but containing 
$\hat{\bm h} = {\bm q}_\sT / |{\bm q}_\sT|$. 

The full expression of the leading-twist differential cross section for the $H_1
H_2^\uparrow \to l^+ l^- X$ process can be written as~\cite{Boer:1999mm}
\bea
\frac{d\sigma}{d\Omega dx_1 dx_2 d{\bm q}_\sT} &= &
\frac{d\sigma^o}{d\Omega dx_1 dx_2 d{\bm q}_\sT} + 
\frac{d\Delta \sigma^\uparrow}{d\Omega dx_1 dx_2 d{\bm q}_\sT}  \nn \\
&= &\frac{\alpha^2}{3Q^2}\,\sum_f\,e_f^2\,\Bigg\{ A(y) \, 
{\cal F}\left[ f_1^f(H_1)\, f_1^f (H_2) \right] \nn \\
& &\mbox{\hspace{2cm}} + B(y) \, \cos 2\phi \, 
{\cal F}\left[ \left( 2 \hat{\bm h}\cdot {\bm p}_{1\sT} \, \hat{\bm h} \cdot 
{\bm p}_{2\sT} - {\bm p}_{1\sT} \cdot {\bm p}_{2\sT} \right) \, 
\frac{h_1^{\perp\,f}(H_1)\,h_1^{\perp\,f}(H_2)}{M_1\,M_2}\,\right] \Bigg\} \nn \\
& &+ \frac{\alpha^2}{3Q^2}\,|{\bm S}_{2\sT}|\,\sum_f\,e_f^2\,\Bigg\{ 
A(y) \, \sin (\phi - \phi_{S_2})\, {\cal F}\left[ \hat{\bm h}\cdot 
{\bm p}_{2\sT} \,\frac{f_1^f(H_1) \, f_{1\sT}^{\perp\,f}(H_2^\uparrow)}{M_2}\right] 
\nn \\
& &\mbox{\hspace{3cm}} - B(y) \, \sin 
(\phi + \phi_{S_2})\, {\cal F}\left[ \hat{\bm h}\cdot {\bm p}_{1\sT} \,
\frac{h_1^{\perp\,f}(H_1) \, h_1^f(H_2^\uparrow)}{M_1}\right] \nn \\
& &\mbox{\hspace{3cm}}  - B(y) \, \sin (3\phi - \phi_{S_2})\, {\cal F}\left[ 
\left( 4 \hat{\bm h}\cdot {\bm p}_{1\sT} \, (\hat{\bm h} \cdot 
{\bm p}_{2\sT})^2 - 2 \hat{\bm h} \cdot {\bm p}_{2\sT} \, {\bm p}_{1\sT} \cdot 
{\bm p}_{2\sT} - \hat{\bm h}\cdot {\bm p}_{1\sT} \, {\bm p}_{2\sT}^2 \right) 
\right. \nn \\
& &\mbox{\hspace{7cm}} \left. \times
\frac{h_1^{\perp\,f}(H_1) \, h_{1\sT}^{\perp\,f}(H_2^\uparrow)}{2 M_1\,M_2^2}\,\right]
\, \Bigg\} \; ,
\label{eq:xsect}
\eea
where $\alpha$ is the fine structure constant, $d\Omega = \sin \theta d\theta
d\phi$, $e_f$ is the charge of the parton with flavor $f$, $\phi_{S_i}$ is
the azimuthal angle of the transverse spin of hadron $i$, and 
\begin{align}
A(y) = \left( \frac{1}{2} - y + y^2 \right) \, \stackrel{\mbox{cm}}{=}\, 
\frac{1}{4}\left( 1 + \cos^2 \theta \right) &\mbox{\hspace{2cm}} 
B(y) = y (1-y) \, \stackrel{\mbox{cm}}{=}\,\frac{1}{4}\, \sin^2 \theta \; . 
\label{eq:lepton}
\end{align}
The TMD functions $f_1^f(H), \, h_1^{\perp\,f}(H)$, describe the distributions 
of unpolarized and transversely polarized partons in an unpolarized hadron $H$,
respectively, while $f_{1\sT}^{\perp\,f}(H^\uparrow)$ and the pair $h_1^f (H^\uparrow), 
h_{1\sT}^{\perp\,f}(H^\uparrow)$, have a similar interpretation but for transversely 
polarized hadrons $H^\uparrow$. The convolutions are defined as 
\be
{\cal F} \left[ DF_1^f(H_1) \, DF_2^f(H_2^{(\uparrow )}) \right] \equiv \int 
d{\bm p}_{1\sT} d{\bm p}_{2\sT}\, \delta \left( {\bm p}_{1\sT} + {\bm p}_{2\sT} - 
{\bm q}_\sT \right) \, \left[ DF_1(x_1,{\bm p}_{1\sT}; \bar{f}/H_1)\, 
DF_2(x_2,{\bm p}_{2\sT}; f/H_2^{(\uparrow )} ) + (f\leftrightarrow \bar{f}) \right] 
\; .
\label{eq:convol}
\ee

In previous papers, we made numerical simulations of the SSA generated in
Eq.~(\ref{eq:xsect}) by the azimuthal dependences $\cos 2\phi$ and 
$\sin (\phi + \phi_{S_2})$ for antiproton beams 
$H_1=\bar{p}$~\cite{Bianconi:2004wu}, by the $\sin (\phi -
\phi_{S_2})$ dependence for proton beams $H_1=p$~\cite{Bianconi:2005yj}, as well 
as for double-polarized Drell-Yan processes with $H_1^\uparrow = H_2^\uparrow =
p^\uparrow$~\cite{Bianconi:2005bd}. A combined measurement of these SSA 
allows to completely determine the intertwined unknown transversity $h_1$ and 
Boer-Mulders function $h_1^\perp$, and the Sivers function $f_{1\sT}^\perp$. The 
Monte Carlo simulation was performed for high-energy proton beams 
($\sqrt{s}=200$ GeV) in the conditions of the foreseen upgrade of RHIC (RHIC 
II), and for antiproton beams of 15 GeV as they could be produced at HESR-GSI. In
the latter case, several scenarios were explored for $5\lesssim \sqrt{s} \lesssim
14$ GeV and $1.5<M<2.5,\, 4<M<9$ GeV, in order to avoid overlaps with the 
strange, charm, and bottom quarkonia [where the elementary annihilation does not 
necessarily proceed through a simple intermediate virtual photon, as it is assumed
in Eq.~(\ref{eq:xsect})]. Here, we reconsider the $\sin (\phi - \phi_{S_2})$ and 
$\sin (\phi + \phi_{S_2})$ asymmetries by using pion beams of 100 GeV as they can 
be produced at COMPASS, in the fixed target mode such as to reach the 
same maximum c.m. energy considered at HESR-GSI, namely $\sqrt{s}\sim 14$ GeV. Most of 
the technical details of the simulation are mutuated from our previous works; hence, 
we will heavily refer to Refs.~\cite{Bianconi:2004wu,Bianconi:2005yj} in the following.

The Monte Carlo events have been generated by the following cross 
section~\cite{Bianconi:2004wu}:
\be
\frac{d\sigma}{d\Omega dx_1 dx_2 d{\bm q}_\sT} = K \, \frac{1}{s}\, 
|{\cal T}({\bm q}_\sT, x_1, x_2, M)|^2 \, \sum_{i=1}^4\, c_i ({\bm q}_\sT, 
x_1,x_2) \, S_i(\theta, \phi, \phi_{_{S_2}}) \; ,
\label{eq:mc-xsect}
\ee
where the event distribution is driven by the elementary unpolarized annihilation,
whose transition amplitude ${\cal T}$ has been highlighted. In
Eq.~(\ref{eq:xsect}), we assume a factorized transverse-momentum dependence in 
each parton distribution such as to break the convolution ${\cal F}$, leading to
\be
|{\cal T}|^2 \approx A(q_\sT,x_1,x_2,M) \, F(x_1,x_2) \; ,
\label{eq:factorized}
\ee
where $q_\sT \equiv |{\bm q}_\sT|$. The function $A$ is parametrized and
normalized as in Ref.~\cite{Conway:1989fs}, where high-energy Drell-Yan $\pi - p$ 
collisions were considered. The average transverse momentum turns out to be 
$\langle q_\sT \rangle > 1$ GeV/$c$ (see also the more recent 
Ref.~\cite{Towell:2001nh}), which effectively reproduces the influence of sizable
QCD corrections beyond the parton model picture of Eq.~(\ref{eq:xsect}). It is
well known~\cite{Altarelli:1979ub} that such corrections induce also large $K$
factors and an $M$ scale dependence in parton distributions, determining their
evolution. As in our previous 
works~\cite{Bianconi:2004wu,Bianconi:2005yj,Bianconi:2005bd}, we conventionally 
assume in Eq.~(\ref{eq:mc-xsect}) that $K=2.5$, but we stress that in an 
azimuthal asymmetry the corrections to the cross sections in the numerator and in 
the denominator should compensate each other, as it turns out to actually happen
at RHIC c.m. square energies~\cite{Martin:1998rz}. Since the range of $M$ values 
here explored is close to the one of Ref.~\cite{Conway:1989fs}, where the 
parametrization of $A, F,$ and $c_i$ in Eq.~(\ref{eq:mc-xsect}) was deduced 
assuming $M$-independent parton distributions, we keep our same previous
approach~\cite{Bianconi:2004wu,Bianconi:2005yj,Bianconi:2005bd} and use
\be
F(x_1,x_2) = \frac{\alpha^2}{12 Q^2}\,\sum_f\,e_f^2\,
f_1^f(x_1; \bar{f}/H_1) \, f_1^f (x_2; f/H_2) + (\bar{f} \leftrightarrow f) \; , 
\label{eq:mcF}
\ee
where the unpolarized distribution $f_1^f (x)$ for various flavors $f=u,d,s,$ is 
taken again from Ref.~\cite{Conway:1989fs}.

The whole solid angle $(\theta, \phi)$ of the final lepton pair in the 
Collins-Soper frame is randomly distributed in each variable. The explicit form 
for sorting it in the Monte-Carlo is~\cite{Bianconi:2004wu,Bianconi:2005yj}
\bea
\sum_{i=1}^4\, c_i (q_\sT,x_1,x_2) \, S_i(\theta, \phi, \phi_{S_2}) &= 
&1 + \cos^2 \theta + \frac{\nu (x_1,x_2,q_\sT)}{2}\, \sin^2\theta \, \cos 2\phi 
\nn \\
& &+ |{\bm S}_{2\sT}|\, c_4 (q_\sT,x_1,x_2)\, S_4 (\theta, \phi, \phi_{S_2}) \; .
\label{eq:mcS}
\eea
If quarks were massless, the virtual photon would be only transversely polarized 
and the angular dependence would be described by the functions $c_1 = S_1 = 1$ 
and $c_2 = 1, \, S_2 = \cos^2 \theta$. Violations of such azimuthal symmetry 
induced by the function $c_3 \equiv \textstyle{\frac{\nu}{2}}$ are due to the 
longitudinal polarization of the virtual photon and to the fact that quarks have 
an intrinsic transverse momentum distribution, leading to the explicit violation
of the socalled Lam-Tung sum rule~\cite{Conway:1989fs}. QCD corrections influence 
$\nu$, which in principle depends also on $M^2$~\cite{Conway:1989fs}. Azimuthal 
$\cos 2\phi$ asymmetries induced by $\nu$ were simulated in 
Ref.~\cite{Bianconi:2004wu} using the simple parametrization of 
Ref.~\cite{Boer:1999mm} and testing it against the previous measurement of
Ref.~\cite{Conway:1989fs}. 

If we consider the Sivers effect in Eq.~(\ref{eq:xsect}), the last term in 
Eq.~(\ref{eq:mcS}) becomes 
\be
S_4 (\theta, \phi, \phi_{S_2}) = (1+\cos^2 \theta) \, \sin (\phi - \phi_{S_2}) 
\label{eq:mcS4-sivers}
\ee
and the corresponding coefficient $c_4$ reads
\be
c_4 (q_\sT,x_1,x_2) = \frac{\sum_f\,e_f^2\,{\cal F}\left[ \hat{\bm h}\cdot 
{\bm p}_{2\sT} \, \displaystyle{\frac{f_1^f(x_1,{\bm p}_{1\sT})\, 
                         f_{1T}^{\perp\, f}(x_2,{\bm p}_{2\sT})}{M_2}} \right]}
{\sum_f\,e_f^2\,{\cal F}\left[ f_1^f(x_1,{\bm p}_{1\sT})\, 
 f_1^f(x_2,{\bm p}_{2\sT}) \right]} \; ,
\label{eq:mcc4-sivers}
\ee
where the complete dependence of the involved TMD parton distributions has been
made explicit. 

Viceversa, if we consider the Boer-Mulders effect in Eq.~(\ref{eq:xsect}) the 
last term in Eq.~(\ref{eq:mcS}) becomes 
\be
S_4 (\theta, \phi, \phi_{S_2}) = \sin^2 \theta \, \sin (\phi + \phi_{S_2}) 
\label{eq:mcS4-boer}
\ee
and the corresponding coefficient $c_4$ reads
\be
c_4 (q_\sT,x_1,x_2) = - \frac{\sum_f\,e_f^2\,{\cal F}\left[ \hat{\bm h}\cdot 
{\bm p}_{1\sT} \, \displaystyle{\frac{h_1^{\perp\, f}(x_1,{\bm p}_{1\sT})\, 
                         h_1^f(x_2,{\bm p}_{2\sT})}{M_1}} \right]}
{\sum_f\,e_f^2\,{\cal F}\left[ f_1^f(x_1,{\bm p}_{1\sT})\, 
 f_1^f(x_2,{\bm p}_{2\sT}) \right]} \; .
\label{eq:mcc4-boer}
\ee

In the following, we will discuss different inputs for the $x$ and ${\bm p}_\sT$ 
dependence of these distributions which allow to calculate the convolutions and 
determine $c_4$. In any case, following 
Refs.~\cite{Bianconi:2004wu,Bianconi:2005yj,Bianconi:2005bd}, the general strategy
is to divide the event sample in two groups, one for positive values "$U$" of 
$S_4$ in Eq.~(\ref{eq:mcS4-sivers}) or (\ref{eq:mcS4-boer}), and another one for 
negative values "$D$", then taking the ratio $(U-D)/(U+D)$. Data are accumulated
only in the $x_2$ bin, i.e. they are summed upon $x_1, \theta,$ and 
$q_\sT$. Statistical errors for the spin asymmetry $(U-D)/(U+D)$ are obtained by 
making 10 independent repetitions of the simulation for each individual case, and 
then calculating for each $x_2$ bin the average asymmetry value and the variance. 
We checked that 10 repetitions are a reasonable threshold to have stable numbers, 
since the results do not change significantly when increasing the number of 
repetitions beyond 6.


\subsection{The Sivers effect}
\label{sec:sivers}

Recently, the HERMES collaboration released new SSA data for the SIDIS 
process on transversely polarized protons~\cite{Diefenthaler:2005gx}, which 
substantially increase the precision of the previous data 
set~\cite{Airapetian:2004tw}. As a consequence, different parametrizations of the
Sivers function $f_{1T}^\perp$ have been extracted from this data set and found 
compatible also with the recent COMPASS data~\cite{Alexakhin:2005iw} (for a useful
comparison among the various approaches see Ref.~\cite{Anselmino:2005an}).
Following Ref.~\cite{Bianconi:2005yj}, we first simulate the Sivers effect using
the parametrization of Ref.~\cite{Anselmino:2005ea}, 
\bea
f_{1T}^{\perp\, f}(x,{\bm p}_\sT) &= &-2\, N_f\,
\frac{(a_f+b_f)^{a_f+b_f}}{a_f^{a_f}\,b_f^{b_f}}\,
x^{a_f}\,(1-x)^{b_f}\,\frac{M_2 M_0}{{\bm p}_\sT^2+M_0^2}\,
f_1^f(x,{\bm p}_\sT) \nn \\
&= &-2\, N_f\,\frac{1}{\pi \, \langle p_\sT^2 \rangle}\,
\frac{(a_f+b_f)^{a_f+b_f}}{a_f^{a_f}\, b_f^{b_f}} \, x^{a_f}\, (1-x)^{b_f}\, 
\frac{M_2 M_0}{{\bm p}_\sT^2+M_0^2}\, e^{-p_\sT^2/\langle p_\sT^2 \rangle}\, 
f_1^f(x) \; ,
\label{eq:pTanselm}
\eea
where $M_2$ is the mass of the polarized proton, $p_\sT \equiv |{\bm p}_\sT|$, 
and $\langle p_\sT^2 \rangle = 0.25$ (GeV/$c$)$^2$ is deduced by assuming a 
Gaussian ansatz for the ${\bm p}_\sT$ dependence of $f_1$ in order to reproduce the 
azimuthal angular dependence of the SIDIS unpolarized cross section (Cahn effect). 
Flavor-dependent normalization and parameters in the $x$ dependence are fitted to 
SIDIS SSA data neglecting the (small) contribution of antiquarks. The resulting 
parameters $M_0$ and $N_f,a_f,b_f,$ with $f=u,d$, are listed in 
Tab.~\ref{tab:pTanselm}. The sometimes poor resolution of the fit forced us to select 
only the central values in order to produce meaningful numerical simulations.

Following the steps described in Sec.~III-1 of Ref.~\cite{Bianconi:2005yj}, in
particular the predicted sign change of $f_{1\sT}^\perp$ when going from SIDIS to
Drell-Yan, we insert the opposite of Eq.~(\ref{eq:pTanselm}) into 
Eq.~(\ref{eq:mcc4-sivers}) and simplify it down to 
\be
c_4 \approx \frac{4 M_0\,q_\sT}{q_\sT^2+4 M_0^2}\, 
\frac{1}{9}\, \left[ 8\, N_u\, \frac{(a_u+b_u)^{a_u+b_u}}{a_u^{a_u}\,b_u^{b_u}} 
\, x_2^{a_u}\,(1-x_2)^{b_u}\, + \, N_d\, 
\frac{(a_d+b_d)^{a_d+b_d}}{a_d^{a_d}\,b_d^{b_d}} \, x_2^{a_d}\,(1-x_2)^{b_d} 
\right] \; .
\label{eq:c4mc-anselm}
\ee

\begin{table}[h]
\caption{\label{tab:pTanselm} Parameters for the Sivers distribution from
Ref.~\protect{\cite{Anselmino:2005ea}}}
\begin{ruledtabular}
\begin{tabular}{cccc}
quark up & {} & quark down & {} \\
\hline
$N_u$ & $0.32 \pm 0.11$ & $N_d$ & $-1.0 \pm 0.12$  \\
$a_u$ & $0.29 \pm 0.35$ & $a_d$ & $1.16 \pm 0.47$ \\
$b_u$ & $0.53 \pm 3.58$ & $b_d$ & $3.77 \pm 2.59$ \\
\hline
$M_0^2$ & $0.32 \pm 0.25$ (GeV/$c$)$^2$ & &  \\
\end{tabular}
\end{ruledtabular}
\end{table}

As an alternative choice, we adopt the new parametrization described in
Ref.~\cite{Bianconi:2005yj}. It is inspired to the one of 
Ref.~\cite{Vogelsang:2005cs}, where the transverse momentum of the detected pion in 
the SIDIS process was assumed to come entirely from the ${\bm p}_\sT$ dependence of 
the Sivers function, and was further integrated out building the fit in terms of
specific moments of the function itself. The $x$ dependence of that approach is
retained, but a different flavor-dependent normalization and an explicit 
${\bm p}_\sT$ dependence are introduced that are bound to the shape of the recent 
RHIC data on $pp^\uparrow \to \pi X$ at $\sqrt{s}=200$ GeV~\cite{Adler:2005in}, where
large persisting asymmetries are found that could be partly due to the leading-twist
Sivers mechanism. The expression adopted is
\bea
f_{1T}^{\perp\, f}(x,{\bm p}_\sT) &=&N_f\,x\,(1-x)\,
\frac{M_2 p_0^2 p_\sT}{(p_\sT^2+\frac{p_0^2}{4})^2}\,f_1^f(x,{\bm p}_\sT) \nn \\
&= & N_f\,x\,(1-x)\,\frac{M_2 p_0^2 p_\sT}{(p_\sT^2+\frac{p_0^2}{4})^2}\,
\frac{1}{\pi \, \langle p_\sT^2 \rangle}\, e^{-p_\sT^2/\langle p_\sT^2 \rangle}\, 
f_1^f(x) \; ,
\label{eq:pTnoi}
\eea
where $p_0 = 2$ GeV/$c$, and $N_u = - N_d = 0.7$. The sign, positive for $u$
quarks and negative for the $d$ ones, already takes into account the predicted sign
change of $f_{1\sT}^\perp$ from Drell-Yan to SIDIS. 

Again, following the steps described in Sec.~III-2 of Ref.~\cite{Bianconi:2005yj}, we 
can directly insert Eq.~(\ref{eq:pTnoi}) into Eq.~(\ref{eq:mcc4-sivers}) and get
\be
c_4 \approx x_2 \, (1-x_2)\, 
\left( \frac{2\, p_0 \, q_\sT}{q_\sT^2+p_0^2} \right)^2 \, \frac{8\, N_u + N_d}{9} 
\; .
\label{eq:c4mc-noi}
\ee
The $q_\sT$ shape is different from Eq.~(\ref{eq:c4mc-anselm}) and the peak position 
is shifted at larger values. This is in agreement with a similar analysis of the 
azimuthal asymmetry of the unpolarized Drell-Yan data (the violation of the Lam-Tung 
sum rule~\cite{Boer:1999mm}). But, more specifically, it is induced by the observed 
$x_{_F}-q_\sT$ correlation in the above mentioned RHIC data for 
$pp^\uparrow \to \pi X$, when it is assumed that the SSA is entirely due to the
Sivers mechanism. This suggests that the maximum asymmetry is reached in 
the upper valence region such that $x_{_F} \approx x_2 \sim \langle q_\sT \rangle /
5$~\cite{Adler:2005in}.


\subsection{The Boer-Mulders effect}
\label{sec:boer}


Contrary to the Sivers effect, the lack of data for the Boer-Mulders effect 
does not allow to build reasonable parametrizations either of 
$h_1^{\perp\,f}(x,{\bm p}_\sT)$ or of $h_1^f(x,{\bm p}_\sT)$. Therefore, similarly to 
what was done in our previous papers~\cite{Bianconi:2004wu,Bianconi:2005bd}, the 
strategy of the numerical simulation is based on making guesses for the input $x$ and 
${\bm p}_\sT$ dependence of the parton distributions, and on trying to determine the
minimum number of events required to discriminate various SSA produced by very
different input guesses. In fact, this would be equivalent to state that in this
case some analytic information on the structure of these TMD parton distributions
could be extracted from the SSA measurement. 

Following the steps in Sec.~IV~C of Ref.~\cite{Bianconi:2004wu} and in Sec.~VI of
Ref.~\cite{Boer:1999mm}, the ${\bm p}_\sT$ dependence of the parton distributions is
parametrized as 
\bea
f_1^f(x,{\bm p}_\sT) &= &\frac{\alpha_\sT}{\pi}\,e^{-\alpha_\sT\,{\bm p}_\sT^2}\,
f_1^f (x) \nn \\
h_1^{\perp\,f}(x,{\bm p}_\sT) &= &\frac{M_{_C}}{{\bm p}_\sT^2 +M_{_C}^2} \, 
f_1^f (x,{\bm p}_\sT) \nn \\
h_1^f(x,{\bm p}_\sT) &= &\frac{\alpha_\sT}{\pi}\,e^{-\alpha_\sT\,{\bm p}_\sT^2}\,
h_1^f (x) \; ,
\label{eq:boerparams}
\eea
where $\alpha_\sT = 1$ GeV$^{-2}$ and $M_{_C}=2.3$ GeV. In particular, the 
${\bm p}_\sT$ dependence of $h_1^{\perp}$ is fitted to the measured $\cos 2\phi$
asymmetry of the corresponding unpolarized Drell-Yan cross section, which is small for
$1\lesssim q_\sT \lesssim 3$ GeV/$c$ (see, for example, Fig.4 in 
Ref.~\cite{Boer:1999mm}). Correspondingly, the 
$\sin (\phi +\phi_S)$ SSA will turn out to be small for the considered statistically
relevant $q_\sT$ range (see Sec.~\ref{sec:out-boer}).  

Inserting the expressions~(\ref{eq:boerparams}) into Eq.~(\ref{eq:mcc4-boer}), we get 
\bea
c_4 &= &-\frac{2 M_{_C}\,q_\sT}{q_\sT^2+4 M_{_C}^2}\, 
\frac{\sum_f\,e_f^2\,f_1^f(x_1; \bar{f}/H_1)\,h_1^f(x_2; f/H_2^\uparrow)+ 
      (\bar{f} \leftrightarrow f)}
     {\sum_f\,e_f^2\,f_1^f(x_1; \bar{f}/H_1)\,f_1^f(x_2; f/H_2)+ 
      (\bar{f} \leftrightarrow f)} \nn \\
&\approx &-\frac{2 M_{_C}\,q_\sT}{q_\sT^2+4 M_{_C}^2} \, 
\frac{f(x_1; \langle \bar{f} \rangle/H_1) \, h_1(x_2;\langle f \rangle/H_2^\uparrow)}
     {f(x_1; \langle \bar{f} \rangle/H_1) \, f_1(x_2;\langle f \rangle/H_2)}  \equiv 
- \frac{2 M_{_C}\,q_\sT}{q_\sT^2+4 M_{_C}^2} \, 
\frac{h_1(x_2;\langle f \rangle/H_2^\uparrow)}{f_1(x_2;\langle f \rangle/H_2)} \; ,
\label{eq:mcc4-boer2}
\eea
where the second step is justified by assuming that the contribution of each flavor can
be approximated by a corresponding average function~\cite{Bianconi:2004wu}. 

Two choices with opposite features will be selected for the ratio 
$h_1(x_2;\langle f \rangle/H_2^\uparrow) / f_1(x_2;\langle f \rangle/H_2)$, namely the
ascending function $\sqrt{x_2}$ and the descending one $\sqrt{1-x_2}$, that both respect
the Soffer bound. The goal is to determine the minimum number of events (compatible 
with the kinematical setup and cuts) required to produce azimuthal asymmetries that 
can be clearly distinguished like the corresponding originating distributions. We 
identify this as the criterion to establish when information on the analytical 
structure of the involved parton distributions can be extracted from SSA data.

\section{Results of the Monte Carlo simulations}
\label{sec:out}

In this Section, we present results for Monte Carlo simulations of both the Sivers
and the Boer-Mulders effects in the Drell-Yan process $\pi^{\pm} p^\uparrow \to 
\mu^+ \mu^- X$ using input from the previous Sec.~\ref{sec:sivers} and \ref{sec:boer},
respectively. The goal is twofold. On one side, to explore the sensitivity of the 
simulated asymmetry to the different input parametrizations of
Eqs.~(\ref{eq:pTanselm}) and (\ref{eq:pTnoi}), as well as to directly verify, within 
the reached statistical accuracy, the predicted sign change of the Sivers function 
between SIDIS and Drell-Yan~\cite{Collins:2002kn}. On the other side, to make
realistic estimates of the minimum number of events required to extract as detailed
information as possible on the chiral-odd distributions $h_1^\perp$ and $h_1$. 

We consider pion beams with energy of 100 GeV hitting a transversely polarized proton
target such that $\sqrt{s}\sim 14$ GeV, i.e. the same c.m. energy available at HESR at
GSI in the socalled asymmetric collider mode with antiprotons of 15 GeV and protons of
3.3 GeV~\cite{Bianconi:2004wu}. The transversely polarized proton target is obtained
from a $NH_3$ molecule where each $H$ nucleus is fully transversely polarized and the
number of "polarized" collisions is 25\% of the total number of 
collisions~\cite{Bianconi:2004wu}. The muon pair invariant mass is constrained in the
range $4<M<9$ GeV, in order to avoid overlaps with the resonance regions of the 
$\bar{c}c$ and $\bar{b}b$ quarkonium systems. At the same time, the theoretical 
analysis based on the leading-twist cross section~(\ref{eq:xsect}) should be well 
established, since higher-twist effects can be classified according to powers of 
$M_p/M$, where $M_p$ is the proton mass. 

In the Monte Carlo, the events are sorted according to the cross 
section~(\ref{eq:mc-xsect}), supplemented by Eqs.~(\ref{eq:factorized}) and 
(\ref{eq:mcF}). The asymmetry is simulated by Eq.~(\ref{eq:mcS}). In particular, for 
the Sivers effect we use Eqs.~(\ref{eq:mcS4-sivers}) and (\ref{eq:c4mc-anselm}) or
(\ref{eq:c4mc-noi}), according to the input parametrization selected for the Sivers
function. For the Boer-Mulders effect, we use Eqs.~(\ref{eq:mcS4-boer}) and
(\ref{eq:mcc4-boer2}). The events are divided in two groups, one for positive values 
($U$) of $\sin (\phi - \phi_{S_2})$ in Eq.~(\ref{eq:mcS4-sivers}) or of 
$\sin (\phi + \phi_{S_2})$ in Eq.~(\ref{eq:mcS4-boer}), and another one for negative 
values ($D$), and taking the ratio $(U-D)/(U+D)$. Data are accumulated only in the 
$x_2$ bins of the polarized proton, i.e. they are summed over in the $x_1$ bins for 
the pion, in the transverse momentum $q_\sT$ of the muon pair and in their zenithal 
orientation $\theta$.

Proper cuts are applied to the $q_\sT$ distribution according to the different inputs.
As for the Sivers effect, the flavor-independent Lorentzian shape in the ${\bm p}_\sT$ 
dependence of Eq.~(\ref{eq:pTanselm}) produces a maximum asymmetry for $q_\sT \sim 1$ 
GeV/$c$ and a rapid decrease for larger values. Consequently, transverse momenta are
selected in the range $0.5< q_\sT < 2.5$ GeV/$c$, because for larger cutoffs the
asymmetry is diluted. For the case of Eq.~(\ref{eq:pTnoi}), the peak position in $q_\sT$
is shifted at higher values and the cut is modified as $1<q_\sT <3$ GeV/$c$. In this 
way, the ratio between the absolute sizes of the asymmetry and the statistical errors 
is optimized for each choice, while the resulting $\langle q_\sT \rangle \sim 1.8$ 
GeV/$c$ is in fair agreement with the one experimentally explored at 
RHIC~\cite{Adler:2005in}. As for the Boer-Mulders effect, we keep the latter cut 
$1<q_\sT <3$ GeV/$c$. The $\theta$ angular dependence for the Boer-Mulders effect is 
constrained in the range $60^{\rm o}< \theta < 120^{\rm o}$ due to 
Eq.~(\ref{eq:mcS4-boer}), because outside these limits the azimuthal asymmetry is too 
small~\cite{Bianconi:2004wu}. On the contrary, for the Sivers effect there is no need 
to introduce cuts because of the $(1+\cos^2 \theta)$ term in 
Eq.~(\ref{eq:mcS4-sivers})~\cite{Bianconi:2005yj}. 


We have considered different initial samples. The Sivers mechanism is explored
starting from $100\,000$ events with the $\pi^-$ beam and $25\,000$ with the 
$\pi^+$ beam, because the Monte Carlo indicates that the cross section involving 
$\pi^+$ is statistically disfavoured by approximately the factor 
1/4~\cite{Bianconi:2005bv}; in such a way, the two samples can be collected in the same 
time. As for the Boer-Mulders effect, the lacking of any parametrization makes it 
impossible to perform an isospin analysis; hence, we used $50\,000$ events with the 
$\pi^-$ beam. Statistical errors for $(U-D)/(U+D)$ are obtained by 
making 10 independent repetitions of the simulation for each individual case, and 
then calculating for each $x_2$ bin the average asymmetry value and the variance. 
We checked that 10 repetitions are a reasonable threshold to have stable numbers, 
since the results do not change significantly when increasing the number of 
repetitions beyond 6.


\begin{figure}[h]
\centering
\includegraphics[width=9cm]{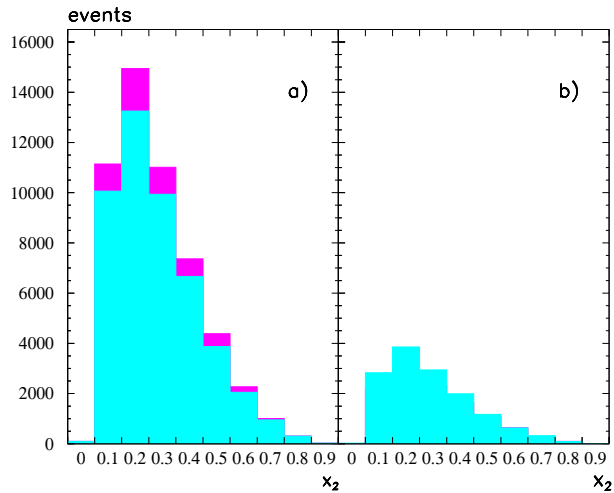}
\caption{The samples of Drell-Yan events for the Sivers effect in the 
$\pi^\pm p^\uparrow \to \mu^+ \mu^- X$ reaction at $\sqrt{s}\sim 14$ GeV, 
$4<M<9$ GeV, and $0.5<q_\sT <2.5$ GeV/$c$, using the parametrization of 
Eq.~(\protect{\ref{eq:pTanselm}}) (see text). a) left panel: $100\,000$ events with 
the $\pi^-$ beam; the darker histogram collects events with positive 
$\sin (\phi - \phi_{S_2})$, the superimposed lighter histogram collects the negative 
ones. b) right panel: the same for $25\,000$ events with the $\pi^+$ beam.}
\label{fig:pi+-anselm_histo}
\end{figure}



\subsection{The Sivers effect}
\label{sec:out-sivers}

In Fig.~\ref{fig:pi+-anselm_histo}, the left panel a) displays the sample of $100\,000$ 
Drell-Yan events for the $\pi^- p^\uparrow \to \mu^+ \mu^- X$ reaction at 
$\sqrt{s} \sim 14$ GeV as they are collected in $x_2$ bins for muon invariant mass in 
the $4<M<9$ GeV range. The right panel b) contains $25\,000$ events for the 
$\pi^+ p^\uparrow \to \mu^+ \mu^- X$ reaction in the same kinematic conditions. Both 
samples can be accumulated approximately in the same time according to 
Eq.~(\ref{eq:c4mc-anselm}) based on the parametrization~(\ref{eq:pTanselm}) of the 
Sivers function~\cite{Anselmino:2005ea}; as already discussed, the transverse momentum 
distribution is constrained in the range $0.5<q_\sT <2.5$ GeV/$c$. For each bin two 
groups of events are stored, one corresponding to positive values of 
$\sin (\phi - \phi_{S_2})$ in Eq.~(\ref{eq:mcS4-sivers}) (represented by the darker 
histogram), and one for negative values (superimposed lighter histogram). Since the
$\bar{q} q \to \gamma^\ast$ mechanism tends to populate the phase space for the lowest
possible $\tau$ values~\cite{Bianconi:2004wu,Bianconi:2005bd,Bianconi:2005yj} compatible
with the explored range $0.08<\tau = x_1 \, x_2<0.4$, this reflects in a
$x_1-$integrated distribution which is peaked for $x_2$ values in the valence domain.



\begin{figure}[h]
\centering
\includegraphics[width=9cm]{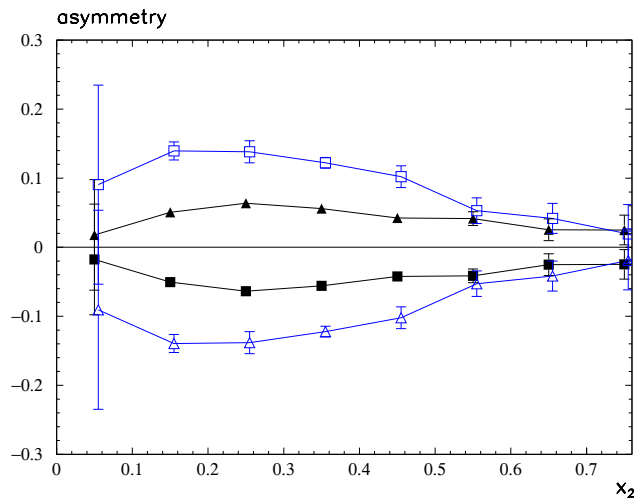}
\caption{The asymmetry $(U-D)/(U+D)$ corresponding to the histograms of
Fig.~\protect{\ref{fig:pi+-anselm_histo}}, where $U$ identifies the darker histograms 
and $D$ the superimposed lighter ones (see text). Triangles for the 
parametrization of Eq.~(\protect{\ref{eq:pTanselm}}) using the $\pi^-$ beam and with 
$N_u>0$; squares for $N_u<0$. Open triangles using the $\pi^+$ beam and with 
$N_u>0$; open squares for $N_u<0$.}
\label{fig:pi+-anselm_as}
\end{figure}


In Fig.~\ref{fig:pi+-anselm_as}, the asymmetry $(U-D)/(U+D)$ is shown for each bin $x_2$
between the events of the previous figure accumulated for the positive $(U)$ and 
negative $(D)$ values of $\sin (\phi - \phi_{S_2})$ in Eq.~(\ref{eq:mcS4-sivers}). 
Average asymmetries and (statistical) error bars are obtained by 10 
independent repetitions of the simulation. Boundary values of $x_2$ beyond 0.7 are 
excluded because of very low statistics. The triangles indicate the results with 
the $\pi^-$ beam obtained by Eq.~(\ref{eq:c4mc-anselm}) assuming that 
$f_{1\sT}^\perp$ changes sign from the parametrization~(\ref{eq:pTanselm}) of the SIDIS 
data to the considered Drell-Yan~\cite{Collins:2002kn}. For sake of comparison, the 
squares illustrate the opposite results that one would obtain by ignoring such 
prediction. Finally, the open triangles and open squares refer to the same situation, 
respectively, but for the $\pi^+$ beam. The sensitivity of the parameters in 
Tab.~\ref{tab:pTanselm} to the HERMES results for the Sivers effect, reflects in a more 
important relative weight of the $d$ quark over the $u$ one in the valence $x_2$ range, 
with opposite signs for the corresponding normalization $N_f, \, f=u,d$. Consequently, 
in the valence picture of the $(\pi^-)\pi^+-p$ collision where the $(\bar{u}u) \, 
\bar{d}d$ annihilation dominates, the SSA for the Drell-Yan process induced by $\pi^+$ 
has opposite sign with respect to $\pi^-$. Moreover, it has an absolute bigger size 
because the $\bar{d}d$ annihilations are weighted more than the $\bar{u}u$ ones. Apart 
for very low $x_2$ values where the parton picture leading to Eq.~(\ref{eq:xsect}) 
becomes questionable, the error bars are very small and allow for a clean 
reconstruction of the asymmetry shape and, more importantly, for a conclusive test of 
the predicted sign change in $f_{1\sT}^\perp$.


\begin{figure}[h]
\centering
\includegraphics[width=9cm]{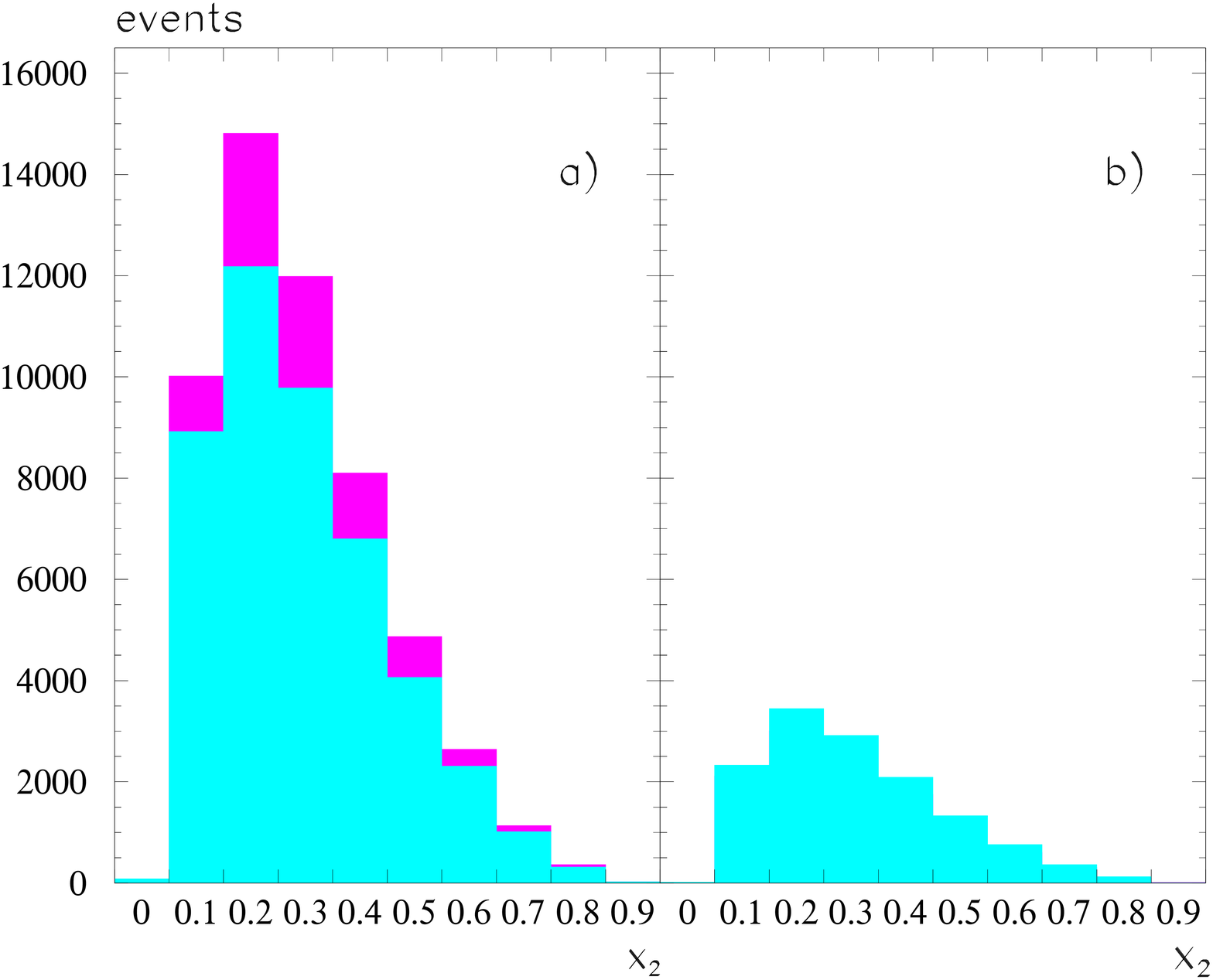}
\caption{The same situation with the same notations as in
Fig.~\protect{\ref{fig:pi+-anselm_histo}}, but for the parametrization of 
Eq.~(\protect{\ref{eq:pTnoi}}) with $1<q_\sT <3$ GeV/$c$ (see text).}
\label{fig:pi+-noi_histo}
\end{figure}


In Fig.~\ref{fig:pi+-noi_histo}, the Drell-Yan events are shown in the same conditions 
and notations as in Fig.~\ref{fig:pi+-anselm_histo}, i.e. in the left panel a)
$100\,000$ events for the $\pi^- p^\uparrow \to \mu^+ \mu^- X$ reaction at $\sqrt{s}
\sim 14$ GeV and for $4<M<9$ GeV, and in the right panel b) $25\,000$ events for the
$\pi^+ p^\uparrow \to \mu^+ \mu^- X$ reaction in the same kinematic conditions. The
difference is that the events are now collected according to Eq.~(\ref{eq:c4mc-noi}) 
based on the parametrization~(\ref{eq:pTnoi}) of the Sivers 
function~\cite{Bianconi:2005yj}; the cut in the transverse momentum distribution is 
now $1<q_\sT <3$ GeV/$c$. Again, the darker histogram refers to events with positive 
$\sin (\phi - \phi_{S_2})$ in Eq.~(\ref{eq:mcS4-sivers}), while the superimposed lighter
histogram to the negative ones. Similarly, the density of events is peaked for $x_2$ 
values in the valence domain because of the dominance of the low $\tau$ portion of the
phase space. 


\begin{figure}[h]
\centering
\includegraphics[width=9cm]{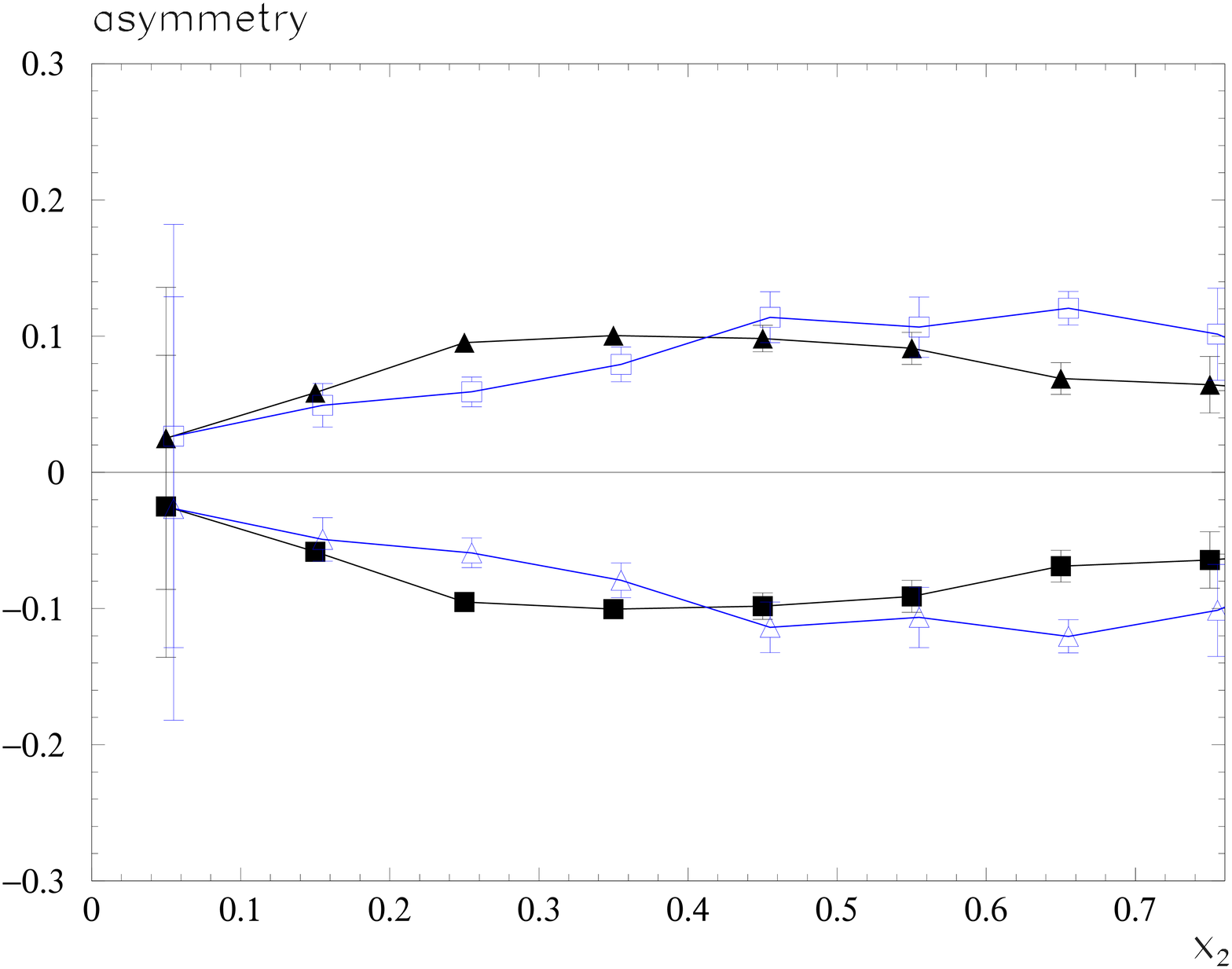}
\caption{The same situation with the same notations as in
Fig.~\protect{\ref{fig:pi+-anselm_as}}, but for the parametrization of 
Eq.~(\protect{\ref{eq:pTnoi}}) (see text).}
\label{fig:pi+-noi_as}
\end{figure}


In Fig.~\ref{fig:pi+-noi_as}, the asymmetry $(U-D)/(U+D)$ is shown for each bin $x_2$
between the events of Fig.~\ref{fig:pi+-noi_histo} accumulated for the positive $(U)$ 
and negative $(D)$ values of $\sin (\phi - \phi_{S_2})$ in Eq.~(\ref{eq:mcS4-sivers}). 
Notations are as in Fig.~\ref{fig:pi+-anselm_as}: the triangles indicate the 
results with the $\pi^-$ beam obtained by Eq.~(\ref{eq:c4mc-noi}) using a positive 
normalization $N_u$, which already accounts for the sign change of $f_{1\sT}^\perp$ 
from SIDIS to Drell-Yan; the squares illustrate the results obtained by ignoring such 
prescription, while the open triangles and open squares refer to the same situation, 
respectively, but for the $\pi^+$ beam. Again, the opposite normalizations of the two
flavors $u,d,$ determine the opposite SSA between the $\pi^-$ and the $\pi^+$ beams. But
now in Eq.~(\ref{eq:pTnoi}) the relative weight of $u$ and $d$ distributions is the
same, hence the absolute sizes of the SSA are approximately the same irrespectively of 
the charge of the $\pi$ beam. As already anticipated in Sec.~\ref{sec:sivers}, the 
$q_\sT$ distribution induced by the parametrization~(\ref{eq:pTnoi}) is also related to 
the observed $x_{_F}-q_\sT$ correlation in the RHIC data for 
$pp^\uparrow \to \pi X$~\cite{Adler:2005in}, when it is assumed that the SSA is 
entirely due to the Sivers mechanism. This suggests that the maximum asymmetry is 
reached in the upper valence region such that $x_{_F} \approx x_2 \sim \langle q_\sT 
\rangle / 5 \sim 0.4$ for the considered cut in $q_\sT$, as it is confirmed in
Fig.~\ref{fig:pi+-noi_as}. Similarly to the case of the other parametrization, the 
statistical error bars are very small and allow for a detailed analysis of the 
(universality) properties of $f_{1\sT}^\perp$.


\begin{figure}[h]
\centering
\includegraphics[width=9cm]{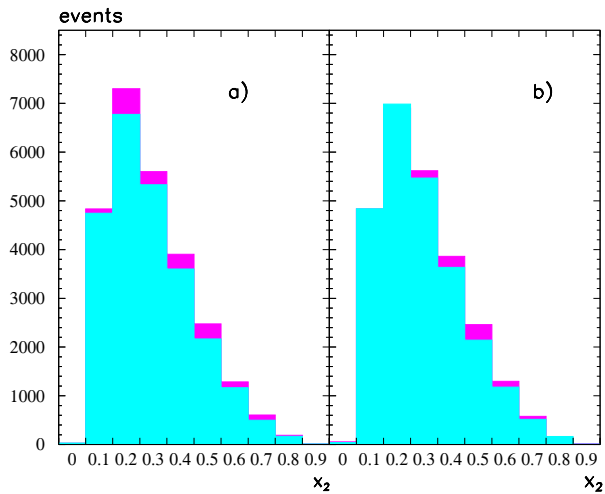}
\caption{The sample of $50\,000$ Drell-Yan events for the Boer-Mulders effect in the 
$\pi^- p^\uparrow \to \mu^+ \mu^- X$ reaction at $\sqrt{s}\sim 14$ GeV, 
$4<M<9$ GeV, and $1<q_\sT <3$ GeV/$c$ (see text). a) left panel for the choice
$h_1(x_2, \langle f \rangle / H_2^\uparrow) / f_1 (x_2, \langle f \rangle / H_2) = 
\sqrt{1-x_2}$ ($\langle f \rangle$ represents a common average term that replaces each
contribution in the flavor sum, for further details see text); the darker histogram 
collects events with positive $\sin (\phi + \phi_{S_2})$, the superimposed lighter 
histogram collects the negative ones. b) right panel: the same for 
$h_1(x_2, \langle f \rangle / H_2^\uparrow) / f_1 (x_2, \langle f \rangle / H_2) = 
\sqrt{x_2}$.}
\label{fig:pi-boer_histo}
\end{figure}



\subsection{The Boer-Mulders effect}
\label{sec:out-boer}

In Fig.~\ref{fig:pi-boer_histo}, a sample of $50\,000$ Drell-Yan events for the 
$\pi^- p^\uparrow \to \mu^+ \mu^- X$ reaction at $\sqrt{s} \sim 14$ GeV is displayed 
in $x_2$ bins for muon invariant mass in the $4<M<9$ GeV range and for 
$1<q_\sT <3$ GeV/$c$. Events are produced by the Boer-Mulders effect contained in
Eq.~(\ref{eq:mcc4-boer2}), where the left panel a) refers to the choice 
$h_1(x_2, \langle f \rangle / H_2^\uparrow) / f_1 (x_2, \langle f \rangle / H_2) = 
\sqrt{1-x_2}$ and the right panel b) to the 
$h_1(x_2, \langle f \rangle / H_2^\uparrow) / f_1 (x_2, \langle f \rangle / H_2) = 
\sqrt{x_2}$ one. Here, $\langle f \rangle$ means that each term contributing to the sum
upon flavors is replaced by a common flavor-averaged parton distribution. Following
previous notations, for each bin the darker histogram represents events with positive 
values of $\sin (\phi + \phi_{S_2})$ in Eq.~(\ref{eq:mcS4-boer}) and the superimposed
lighter histogram indicates the ones with negative values. Similarly, the density of 
events is peaked for $x_2$ values in the valence domain because of the dominance of the 
low $\tau$ portion of the phase space.



\begin{figure}[h]
\centering
\includegraphics[width=9cm]{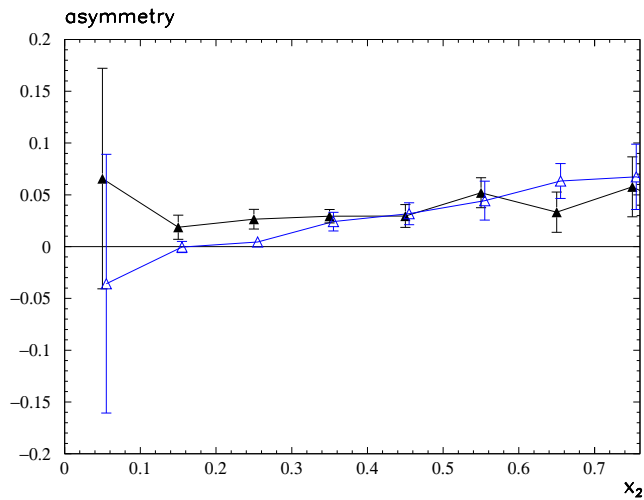}
\caption{The asymmetry $(U-D)/(U+D)$ corresponding to the histograms of
Fig.~\protect{\ref{fig:pi-boer_histo}}, where $U$ identifies the darker histograms 
and $D$ the superimposed lighter ones (see text). Triangles for 
$h_1(x_2, \langle f \rangle / H_2^\uparrow) / f_1 (x_2, \langle f \rangle / H_2) = 
\sqrt{1-x_2}$. Open triangles for 
$h_1(x_2, \langle f \rangle / H_2^\uparrow) / f_1 (x_2, \langle f \rangle / H_2) = 
\sqrt{x_2}$.}
\label{fig:pi-boer_as}
\end{figure}



In Fig.~\ref{fig:pi-boer_as}, the asymmetry $(U-D)/(U+D)$ is shown for each bin $x_2$
between the events of Fig.~\ref{fig:pi-boer_histo} accumulated for the positive $(U)$ 
and negative $(D)$ values of $\sin (\phi + \phi_{S_2})$ in Eq.~(\ref{eq:mcS4-boer}).
Triangles correspond to the $\sqrt{1-x_2}$ input function, open triangles to the
$\sqrt{x_2}$ one. Both choices respect the Soffer bound between $h_1$ and $f_1$ and have
an overall normalization 2/3, which seems a reasonable expectation on the basis of
lattice results and first SIDIS experimental data~\cite{Boer:1999mm}. The error bars 
represent statistical errors only. As it is evident in the figure, the open triangles
describe a SSA which statistically reflects the ascending trend of the input function
$\sqrt{x_2}$, while it is not the case for the other choice. Despite the small error
bars, which allow to state that both SSA are nonvanishing and to distinguish the two 
cases in the narrow range $0.1<x_2<0.3$, it is not possible to conclude that specific
information on the analytic dependence of $h_1(x)$ can be extracted from such
simulation, contrary to what is claimed in Ref.~\cite{Sissakian:2005yp}. The reached 
statistical accuracy indicates that the size of the sample may
not be responsible for such failure. Rather, the ${\bm p}_\sT$ dependence of 
$h_1^{\perp}$ in Eq.~(\ref{eq:boerparams}) induces the overall small size of the
displayed SSA. Moreover, the Soffer bound and the flavor independence of the analysis 
further reduce the selectivity power of the final Monte
Carlo output. In particular, the latter issue calls for specific parametrizations
of $h_1^f(x,{\bm p}_\sT)$ and $h_1^{\perp\,f}(x,{\bm p}_\sT)$, whose unavailability 
reflects the lacking of experimental data for $\sin (\phi + \phi_{S_2})$ 
asymmetries in single-polarized Drell-Yan processes.

\section{Conclusions}
\label{sec:end}


In a series of previous papers~\cite{Bianconi:2004wu,Bianconi:2005bd}, we investigated
the spin structure of the proton using numerical simulations of single- and 
double-polarized Drell-Yan Single-Spin Asymmetries (SSA) for the 
$\bar{p}^{(\uparrow )} p^\uparrow \to \mu^+ \mu^- X$ process as well as for the 
$p p^\uparrow \to \mu^+ \mu^- X$ one~\cite{Bianconi:2005yj}. We selected muon pair 
invariant masses in the range $4<M<9$ GeV (and also $12<M<40$ GeV for the case of
proton beams), where there is no overlap with the resonance regions of the $\bar{c}c$ 
and $\bar{b}b$ quarkonium systems and the elementary annihilation can be safely assumed
to proceed through the $\bar{q}q\to \g^\ast$ mechanism. In particular, the Monte Carlo 
was based on the Drell-Yan leading-twist cross section, because higher twists may be 
suppressed as $M_p/M$, where $M_p$ is the proton mass.

As for single-polarized reactions, two interesting contributions generate azimuthal 
asymmetries of the kind $\sin (\phi + \phi_S)$ and $\sin (\phi - \phi_S)$, where $\phi$ 
and $\phi_S$ are the azimuthal orientations of the plane containing the final muon pair
and of the proton polarization, respectively, with respect to the reaction plane. The
first one involves the convolution of the transversity $h_1$, the missing piece 
necessary to complete the knowledge of the nucleon spin structure at leading twist, and 
the Boer-Mulders $h_1^\perp$, another chiral-odd parton density which is most likely 
responsible for the violation of the Lam-Tung sum rule, the long-standing problem of an 
anomalous $\cos 2\phi$ asymmetry of the corresponding unpolarized Drell-Yan cross 
section~\cite{Boer:1999mm}. The second convolution involves the socalled Sivers 
function $f_{1T}^\perp$~\cite{Sivers:1990cc}, a "naive T-odd" partonic density that 
describes how the distribution of unpolarized quarks is distorted by the transverse 
polarization of the parent hadron. As such, $f_{1T}^\perp$ contains unsuppressed 
information on the orbital motion of hidden confined partons and on their spatial 
distribution inside the proton~\cite{Burkardt:2003je}.

In this paper, we have reconsidered the same scenario but for the 
$\pi^\pm p^\uparrow \to \mu^+ \mu^- X$ process at $\sqrt{s}\sim 14$ GeV, that 
can be reached at COMPASS with pion beams of energy 100 GeV and transversely polarized
proton fixed targets. As with antiproton beams, the elementary mechanism is dominated 
by the annihilation between valence partons (from $p$) and valence antipartons (from 
$\pi$). Taking advantage on the high statistics reachable with pions, in our Monte 
Carlo we have simulated both $\sin (\phi \pm \phi_S)$ SSA in the Drell-Yan cross 
section. For the Sivers effect we have used two parametrizations of $f_{1\sT}^\perp$: 
the one of Ref.~\cite{Anselmino:2005ea}, which was deduced by fitting the recent HERMES 
data for the $\sin (\phi - \phi_S)$ SSA in SIDIS~\cite{Diefenthaler:2005gx}; the one of 
Ref.~\cite{Bianconi:2005yj}, which is constrained by the recent RHIC data for 
the $pp^\uparrow \to \pi X$ process at higher energy~\cite{Adler:2005in}, when it is
assumed that the SSA is driven by the Sivers mechanism only. The main difference is 
that the former displays an emphasized relative importance of the unfavoured $d$ quark, 
and it gives an average transverse momentum $\langle q_\sT \rangle$ of the lepton pair 
lower than the latter. Consistently, we have built SSA by integrating the $q_\sT$ 
distribution with adequate cuts, namely $0.5<q_\sT < 2.5$ GeV/$c$ for the former 
parametrization, and $1<q_\sT<3$ GeV/$c$ for the latter one. Results have been 
presented as binned in the parton momenta $x_2$ of the polarized proton, i.e. by 
integrating also upon the antiparton partner momenta $x_1$ and the zenithal muon pair 
distribution $\theta$ with no further cuts. For the Boer-Mulders effect, since there is 
no such abundance of data and fits, we have used, as we did in 
Ref.~\cite{Bianconi:2004wu}, very different input test functions and we have explored 
the sensitivity of the simulated $\sin (\phi + \phi_S)$ asymmetry within the reached 
statistical accuracy, integrating $q_\sT$ in the range $1<q_\sT<3$ GeV/$c$. Again, 
results have been presented as binned in $x_2$ by integrating also upon $x_1$ and 
$\theta$, but with the further constraint $60^{\rm o}<\theta <120^{\rm o}$ induced by
the factor $\sin^2 \theta$ which drives the angular distribution of muon pairs.

Given the very different situations for the two analyses, also the goals are different.
For the Sivers effect, the numerical simulation aims to establish the necessary 
statistical accuracy to distinguish different input parametrizations and to test the 
(universality) properties of the Sivers function, in particular its predicted sign 
change when going from SIDIS to the Drell-Yan process~\cite{Collins:2002kn}. As for the
Boer-Mulders effect, the goal is to make input guesses and to try to determine the
minimum number of events required to discriminate various SSA produced by very
different input guesses, that would allow to extract as detailed information as 
possible on the chiral-odd distributions $h_1^\perp$ and $h_1$.

In all cases, sorted events have been divided in two groups, corresponding to opposite 
azimuthal orientations of the muon pair with respect to the reaction plane 
(conventionally indicated with $U$ and $D$), and the asymmetry $(U-D)/(U+D)$ has been 
considered. Statistical errors have been obtained by making 10 independent repetitions 
of the simulation for each individual case and, then, calculating for each $x_2$ bin 
the average asymmetry and the variance. For the Sivers effect, a starting sample of 
$100\,000$ events has been selected for the $\pi^-$ beam. Since, from the Monte Carlo,
the cross section with $\pi^+$ turns out statistically unfavoured by a factor
1/4~\cite{Bianconi:2005bv}, we have reduced the sample to $25\,000$ events for the 
$\pi^+$ beam in order to compare situations with the same "effective luminosity". 
As for the Boer-Mulders effect, because of the unavailability of fits and isospin 
analyses, we have used $50\,000$ events with the $\pi^-$ beam. In all cases, the 
$1/\tau$ behaviour of the cross section, induced by the $\g^\ast$ propagator, has a 
twofold effect. It produces the highest density of events for bins in the valence 
domain, typically for $x_2 \sim 0.3$. At the given $\sqrt{s}$, it also grants that the 
considered invariant mass range allows to explore the most populated portion of phase 
space, while avoiding overlaps with ranges where the elementary mechanism could be more 
complicated and the leading-twist analysis more questionable. The direct consequence is 
that, with a very large statistics of pions available, very small error bars are 
reached, except for boundary $x_2$ values. 

The availability of different parametrizations of the Sivers function, obtained from
independent sets of data, allows for a deep analysis of the flavor dependence of the
resulting Drell-Yan SSA, as well as for a test of the universal properties of this
parton density. It turns out that the asymmetry always changes sign when switching from
the $\pi^-$ to the $\pi^+$ beam, because in the valence picture of the 
$(\pi^-) \pi^+-p$ collision the $(\bar{u}u) \, \bar{d}d$ annihilation dominates, and
both the parametrizations here considered have weights with opposite signs for the $u$ 
and $d$ valence quarks. The parametrization of 
Ref.~\cite{Anselmino:2005ea}, being deduced by SIDIS data for the Sivers 
effect~\cite{Diefenthaler:2005gx}, displays a more important relative weight of the 
$d$ quark over the $u$, which reflects in a smaller absolute size of the SSA with the 
$\pi^-$ beam with respect to the $\pi^+$ case. No such evidence is shown by the 
parametrization of Ref.~\cite{Bianconi:2005yj}, constrained by data for the 
$pp^\uparrow \to \pi X$ process at $\sqrt{s}=200$ GeV~\cite{Adler:2005in}, where also 
the higher $\langle q_\sT \rangle$ induces a maximum of the asymmetry at higher $x_2$, 
typically $x_2 \sim 0.4$. In both the considered cases, we have simulated the asymmetry 
assuming or neglecting the predicted sign change of the Sivers function when replacing 
the SIDIS with the Drell-Yan process~\cite{Collins:2002kn}. The corresponding results 
have, of course, opposite signs, but, noticeably, the very small statistical error 
bars allow to clearly distinguish between one choice or the other extreme. We conclude 
that with the considered sample of events it is possible to perform such important test 
of nonperturbative QCD using pion beams and transversely polarized proton targets in 
the kinematic conditions that can be prepared at COMPASS. 

Unfortunately, for the Boer-Mulders effect the lack of data and parametrizations of the
involved parton distributions forbids a thorough analysis. The ${\bm p}_\sT$ 
dependence of $h_1^{\perp}$ is inherited by fitting the measured $\cos 2\phi$
asymmetry of the corresponding unpolarized Drell-Yan cross section; for the 
statistically relevant range $1\lesssim q_\sT \lesssim 3$ GeV/$c$, the 
$\sin (\phi +\phi_S)$ asymmetry turns out to be small. We have further approximated 
the transversity distribution by using a "flavor-averaged" ratio 
between $h_1(x_2)$ itself and the unpolarized parton distribution $f_1(x_2)$, 
and we have simulated it by integrating upon $x_1, q_\sT, \theta ,$ and inserting very 
different input test functions of $x_2$, one ascending and one descending, but all 
satisfying the general constraints (like the Soffer bound, that puts a strong upper 
bound on the size of $h_1$). The small 
statistical errors allow to conclude that the resulting $(U-D)/(U+D)$ asymmetries, 
though small, are certainly nonvanishing. But the displayed trends in $x_2$ are very 
similar and do not reflect the very different inputs. Hence, we conclude that in the
present stage a measurement of such an asymmetry would not help in extracting 
information on the transversity $h_1$ and/or the Boer-Mulders function $h_1^\perp$.


\begin{acknowledgments}

This work is part of the European Integrated Infrastructure Initiative in Hadron
Physics project under the contract number RII3-CT-2004-506078.

\end{acknowledgments}


\bibliographystyle{apsrev}
\bibliography{hadron.bib}

\end{document}